\documentclass[10pt,draft]{dis03}
\usepackage{epsf,amsmath}

\begin{document}

\title{Impact of the running QCD coupling on the value of the
intercept of the structure function $g_1$ singlet}

\author{B.I.~Ermolaev\footnote{
Permanent Address: Ioffe Physico-Technical Institute, 194021
 St.Petersburg, Russia}\\
CFTC, University of Lisbon, Portugal\\
M.~Greco\\
University Rome III, Rome, Italy\\
S.I.~Troyan\\
St.Petersburg Institute of Nuclear Physics, Russia}

\maketitle

\begin{abstract}
The running QCD coupling effects
for $g_1$ singlet are studied.
We predict  that
asymptotically $g_1^S \sim x^{- \Delta_S}$, with the singlet intercept
$\Delta_S \approx 0.85$.
\end{abstract}

\section{Introduction}

The flavour nonsinglet $g_1^{NS}$ and singlet $g_1^S$ components
of the spin structure function $g_1$
are presently the objects of intensive theoretical investigations.
For kinematic region of large $x$, both  $g_1^{NS}$ and $g_1^S$
were
calculated first in Refs.~\cite{ar},\cite{ap}.
Since that, the DGLAP approach has become
the standard instrument for the theoretical description of $g_1$.
This approach provides
 good agreement between experimental data and theoretical predictions.
However, despite
this agreement, it is known that from a theoretical point of view
the DGLAP equations, being extrapolated into the region of small $x$,
are not supposed to work
well because they neglect terms  $\sim (\alpha_s \ln^2 (1/x))^n, ~n > 1$ in
a systematical way.
These double-logarithmic (DL) contributions are small
when $x\sim 1$ but  become very
important when $x\ll 1$  and should be
accounted for to all orders in $\alpha_s$. Total resummation of
DL terms was done in  Refs.~\cite{ber1} for $g_1^{NS}$
and in Ref.~\cite{ber} for $g_1^S$. These calculations
proved that $g_1^{NS}$ and $g_1^S$ have the
Regge behaviour $\sim x^{- \Delta}$
when $x\to 0$. However, the QCD coupling in Refs.~\cite{ber1,ber}
is kept fixed.
Therefore the DL intercepts obtained in \cite{ber1,ber}
are proportional to
$\sqrt{\alpha_s}$ fixed at an unknown scale, which makes the
results of Refs.~\cite{ber1,ber} to be unclear for practical
use. Similarly to the DGLAP equations, the DGLAP-like parametrisation
$\alpha_s = \alpha_s(Q^2)$ should not be used for $g_1$ at small $x$
(see  Ref.~\cite{egt1}).
Below we account for the running $\alpha_s$ effects for the description of
the small-$x$ behaviour of the flavour singlet component of $g_1$.
This paper in based on results obtained in Ref.~\cite{egtsingl}.

\section{IREE for the structure function $g_1$}

In order to obtain $g_1$ with all DL contributions accounted for at small
 $x$, we do not use the DGLAP equations. Instead of that,
we construct and solve a set of infrared evolution equations (IREE), i.e.
equations for the evolution  with respect to the infrared cut-off $\mu$ in
the transverse
momentum space. The cut-off $\mu$ is introduced as the starting point of
the evolution with respect to both $Q^2$ and  $x$. In contrast to the
DGLAP equations where only the $Q^2$ -evolution is studied, the IREE are
two-dimensional.
In order to derive these IREE, it is convenient to operate with
the spin-dependent
invariant amplitude $M_q$ corresponding to the
spin-dependent part of the
forward Compton scattering amplitude
$M_{\mu \nu}(s, Q^2)$ related to $g_1$ as follows:
\begin{equation}
\label{g1m}
g_1(s, Q^2) = \frac{1}{\pi} \Im_s M_q(s, Q^2) .
\end{equation}

We have used here the standard
notations $x =  Q^2/ 2pq$ and $Q^2 = -q^2 >0$, with  $q$ being
the momentum of the off-shell photon
and $p$ being the momentum of the (nearly)
on-shell quark.
The subscript $q$ at $M_q$ means that the off-shell photon
is scattered by the quark.
We will need also the spin-dependent
invariant amplitude, $M_g(s, Q^2)$ of the other forward Compton
scattering where the off-shell photon is scattered by a nearly
on-shell gluon with momentum $p$.
It is convenient to use the Sommerfeld-Watson
(SW) transform for the scattering amplitudes. However, the SW transform is
defined for $M_{q,g}$:
\begin{equation}
\label{mellin}
 M_r(s, Q^2) =
\int_{- \imath \infty}^{\imath \infty}\frac{d \omega}{2 \pi \imath}
(s/ \mu^2)^{\omega}
\xi(\omega)  F_r(\omega, Q^2)
\end{equation}
where $r = q, g$ and $\xi(\omega) =
[1 - e^{ - \imath \pi \omega}]/ 2 \approx  \imath \pi \omega / 2$
is the negative signature factor.

In these terms, the system of IREE for  $F_r(\omega, Q^2)$
can be written down as follows:
\begin{eqnarray}
\label{system}
\big( \omega + \frac{\partial}{\partial y}\big) F_q(\omega, y) =
\frac{1}{8 \pi^2} \big[ F_{qq}(\omega) F_q(\omega, y)  +
F_{qg}(\omega) F_g (\omega, y)\big] ~, \nonumber \\
\big( \omega + \frac{\partial}{\partial y}\big) F_g(\omega, y) =
\frac{1}{8 \pi^2} \big[ F_{gq}(\omega) F_q(\omega, y)  + F_{gg}(\omega)
 F_g(\omega, y) \big] ~
\end{eqnarray}
where the anomalous dimensions $F_{ik}$, with $i,k = q,g$,
correspond to the forward amplitudes
for quark and/or gluon scattering, having used the standard
DGLAP notations for the subscripts. It is convenient to
use normalized amplitudes $H_{ik}$~:
$H_{ik}(\omega) = (1/ 8 \pi^2)F_{ik}(\omega) $~,
that obey the following equations:
\begin{eqnarray}
\label{systemh}
\omega H_{qq} &=& a_{qq} +V_{qq} +H_{qq}^2 + H_{qg}H_{gq}, \nonumber \\
\omega H_{gg} &=& a_{gg} +V_{gg} +H_{gg}^2 + H_{gq}H_{qg}, \nonumber \\
\omega H_{qg} &=& a_{qg} +V_{qg} +H_{qg}( H_{qq} + H_{gg}), \nonumber \\
\omega H_{gq} &=& a_{gq} +V_{gq} +H_{gq}( H_{qq} + H_{gg}) ~ .
\end{eqnarray}
with
\begin{equation}
\label{vborn}
V_{ik} =\frac{ m_{ik}}{\pi^2}D(\omega) ,
m_{qq} =  \frac{C_F}{2N},~
m_{gg} = -2N^2,~
m_{qg} = \frac{N}{2},~
m_{gq} = -N C_F.
\end{equation}

Furthermore $D(\omega)$ in Eq.~(\ref{vborn})
accounts for the running QCD effects for
$V_{ik}$,
\begin{equation}
\label{d}
D(\omega) = \frac{1}{2b^2} \int_0^{\infty} d \rho e^{- \omega \rho}
\ln \big( (\rho + \eta)/ \eta\big)
\Big[  \frac{\rho + \eta}{(\rho + \eta)^2 + \pi^2} +
\frac{1}{\rho + \eta}\Big]
\end{equation}
where $\eta = \ln(\mu^2/ \Lambda_{QCD}^2)$ and $b = (33 - 2 n_f)/ 12 \pi$.

All Born terms
$a_{ik}$ in Eqs.~(\ref{systemh}) are parametrized as follows:
\begin{equation}
\label{aborndiag}
a_{qq} = \frac{A(\omega)C_F}{2\pi},\qquad
~a_{gg} = \frac{2A(\omega) N}{\pi},~
\end{equation}
where
\begin{equation}
\label{a}
A(\omega) = \frac{1}{b} \Big[\frac{\eta}{\eta^2 + \pi^2} -
\int_{0}^{\infty} \frac{d \rho e^{- \omega \rho}}
{(\rho + \eta)^2 + \pi^2}\Big] ~,
\end{equation}

\begin{equation}
\label{aborn}
a_{qg} = -\frac{n_f A'(\omega)}{2\pi},\qquad
a_{gq} =\frac{\alpha_s A'(\omega)C_F}{\pi} ~,
\end{equation}
with
\begin{equation}
\label{aprime}
A'(\omega) = \frac{1}{b} \Big[\frac{1}{\eta} -
\int_{0}^{\infty} \frac{d \rho e^{- \omega \rho}}
{(\rho + \eta)^2}\Big] =  \frac{1}{b}
\int_{0}^{\infty} \frac{d \rho e^{- \omega \rho}}
{\rho + \eta} ~.
\end{equation}

The  $\pi^2$-terms take place when the arguments of $\alpha_s$ are
 time-like.
Finally, we arrive at the following
expression for the singlet $g_1$:
\begin{equation}
\label{g1general}
g_1^S(x, Q^2) = \int_{- \imath \infty}^{\imath \infty}
\frac{d \omega}{2 \pi \imath} (1/ x)^{\omega}
\Big[\tilde{C}_+(\omega) e^{\Omega_+ y} +
\tilde{C}_-(\omega) e^{\Omega_- y} \Big]
\end{equation}
where
\begin{equation}
\label{omega}
\Omega_{\pm} =
\left[H_{qq} + H_{gg} \pm
\sqrt{(H_{qq} - H_{gg})^2 + 4H_{qg}H_{gq}} \right] /2 ~.
\end{equation}

\section{Small-$x$ asymptotics of $g_1$}

\begin{figure}[!thb]
\begin{center}
\begin{picture}(270,180)
\put(10,10){
\epsfbox{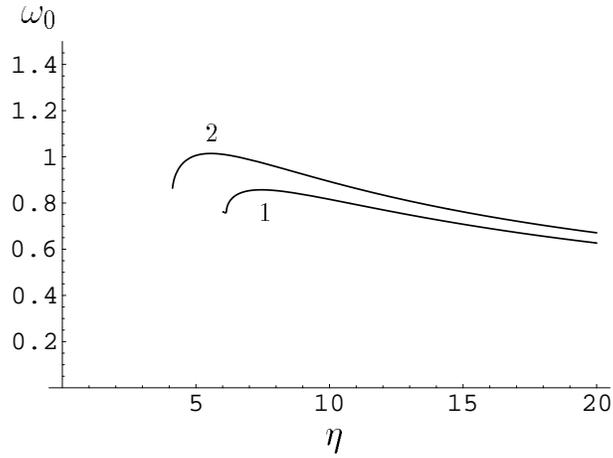}}
\end{picture}
\caption[]{Dependence on $\eta$ of the rightmost root of
Eq.~(\ref{master}), $\omega_0$. Curve~2 corresponds to the case
when gluon contributions only are taken into account;
curve~1 is the result of accounting for both gluon and quark
contributions.}
\label{fig1}
\end{center}
\end{figure}

Eqs.~(\ref{g1general},\ref{omega}) give an explicit expression
for $g_1^S$, which accounts for both  DL contributions and running
$\alpha_s$ effects.
When the limit $x \to 0$ is considered,  one can neglect the second term in
Eq.~(\ref{g1general}) and simplify the expression for $\Omega_+$. Indeed,
the behaviour of $g_1^S$ in this limit is driven by the leading
singularity in Eq.~(\ref{omega}). The singularities of   $\Omega_+$ are
related to the branching points of the square root.
Introducing variables $b_{ik}$ as
\begin{equation}
\label{b}
b_{ik} = a_{ik} + V_{ik} ,
\end{equation}
one obtains that the leading
singularity  of $\Omega_+(\omega)$ is given by the rightmost
root,  $\omega_0$, of the equation below:
\begin{equation}
\label{master}
\omega^4 - 4 (b_{qq} + b_{gg})\omega^2 + 16
(b_{qq} b_{gg} -   b_{qg} b_{gq}) = 0~,
\end{equation}
which predicts that
\begin{equation}
\label{g1asympt}
g_1^S \sim (1/x)^{\omega_0}(Q^2/\mu^2)^{\omega_0/2}
\end{equation}
when $x \to 0$. Eq.~(\ref{master})
 can be solved numerically. In our approach, $\omega_0$ depends on
$\eta = \ln(\mu^2/ \Lambda_{QCD}^2)$. The result of the numerical
calculation for $\omega_0 = \omega_0(\eta)$ is represented by the curve--1
in  Fig.~\ref{fig1}.
%%%%%%%%%%%%%%%%%%%%%%%%%%%%%%
%%%%%%%%%%%%%%%%%%%%%%%%%%%%%%
This curve first grows with $\eta$,
achieves a maximum where approximately $\omega_0 = 0.82$ and smoothly
decreases for large $\eta$. In other
words we have obtained  that the intercept depends strongly on the
infrared cut-off $\mu$ for small values of  $\mu$ and smoothly thereafter.
Quite a similar situation was occurring in Refs.~\cite{egt2} for
the intercept of the non-singlet structure function $g^{NS}_1$.
We suggest a possible  explanation for this effect.  The cut-off $\mu$ is
defined as the starting point in the description of  the
perturbative  evolution. Everything that affects the
intercept at  scales smaller than $\mu$ is attributed to
non-leading effects and/or non-perturbative
contributions. Had they been accounted for, the intercept would have
been $\mu$-independent. Without those non-leading/non-perturbative effects
taken into account, we then observe an important
 $\mu$-dependence, which becomes weaker for large $\mu$. The maximal
perturbative contribution
to the intercept $\Delta_S$ of $g_1$ is obtained from  the
maximal value of $\omega_0$. Therefore
we estimate the intercept as
\begin{equation}
\label{intercept}
\Delta_s = max (\omega_0(\eta)) \approx 0.85 ~.
\end{equation}

Eq.~(\ref{intercept}) includes the contributions of both virtual quarks
and gluons.
These contributions have opposite signs and partly cancel each other.
It is interesting to note that when only
virtual gluon contributions are taken into account,
this purely gluonic
intercept, $\Delta_g$ is given by the maximum of the curve--2 in
Fig.~\ref{fig1}, which is  slightly greater than 1.
This  value
exceeds the unitarity limit, similarly to the intercept
of the LO BFKL, though in a much softer way.
Fortunately, by including also the contributions of the virtual quarks
 the intercept decreases down to $\Delta_S$ of Eq.~(\ref{intercept}),
without violating unitarity.

\section*{Acknowledgement}
We are grateful to S.I.~Krivonos for useful discussions concerning the
numerical calculations.
The work is supported by grants POCTI/FNU/49523/2002
and RSGSS-1124.2003.2 .

\end{document}